\documentclass[9pt,twocolumn,twoside,lineno]{pnas-new}

\templatetype{pnasresearcharticle} 

\title{Data-driven quantitative modeling of bacterial active nematics}

\author[a,b,1]{He Li}
\author[c,d,1]{Xia-qing Shi} 
\author[a,b]{Mingji Huang}
\author[a,b]{Xiao Chen}
\author[e]{Minfeng Xiao}
\author[e]{Chenli Liu}
\author[d,f,2]{Hugues Chat\'e}
\author[a,b,g,2]{H. P. Zhang}

\affil[a]{School of Physics and Astronomy, Shanghai Jiao Tong University, Shanghai 200240, China}
\affil[b]{Institute of Natural Sciences, Shanghai Jiao Tong University, Shanghai 200240, China}
\affil[c]{Center for Soft Condensed Matter Physics and Interdisciplinary Research, Soochow University, Suzhou 215006, China}
\affil[d]{Service de Physique de l'Etat Condens\'e, CEA, CNRS, Universit\'e Paris-Saclay, CEA-Saclay, 91191 Gif-sur-Yvette, France}
\affil[e]{Institute of Synthetic Biology, Shenzhen Institutes of Advanced Technology, Chinese Academy of Sciences, Shenzhen, China}
\affil[f]{Computational Science Research Center, Beijing 100094, China}
\affil[g]{Collaborative Innovation Center of Advanced Microstructures, Nanjing 210093, China
}

\leadauthor{Li} 

\significancestatement{
Active nematics are non-equilibrium fluids consisting of elongated units driven at the individual scale. 
They spontaneously exhibit complex spatiotemporal dynamics, and have attracted the attention of scientists from many disciplines. 
Here, we introduce a novel experimental system (made of filamentous bacteria) and a new type of microscopic model for active nematics. 
Simultaneous measurements of orientation and velocity fields yield comprehensive experimental data 
that can be used to identify optimal values for all important parameters in the model. 
At these optimal parameters, the model quantitatively reproduces all experimentally measured features. 
This, in turn, reveals key processes governing active nematics. 
Our versatile approach successfully combines quantitative experiments and data-driven modeling; it can be used to study other dense active systems.}

\authorcontributions{H.L., X.Q.S., H.C. and H.P.Z. designed research; H.L., X.Q.S., M.H., X.C. M.X., C.L., H.C. and H.P.Z. performed research; H.L., X.Q.S., H.C. and H.P.Z. analyzed data; and H.L., X.Q.S., H.C. and H.P.Z. wrote the paper.}
\authordeclaration{The authors declare no conflict of interest.}
\equalauthors{\textsuperscript{1}H. L. and X.q. S. contributed equally to this work.}
\correspondingauthor{\textsuperscript{2}To whom correspondence should be addressed. E-mail: hugues.chate@cea.fr hepeng\_zhang@sjtu.edu.cn}

\keywords{bacteria collective motion $|$ active nematics $|$ topological defects $|$ quantitative modeling} 

\begin{abstract}
Active matter comprises individual units that convert 
energy into mechanical motion. In many examples, such as bacterial systems
and biofilament assays, constituent units are elongated and can give rise to
local nematic orientational order.
Such `active nematics' systems have attracted much attention from both theorists and experimentalists.
However, despite intense research efforts, data-driven quantitative
modeling has not been achieved, a situation mainly due to the lack of systematic
experimental data and to the large number of parameters of current
models. Here we introduce a new active nematics system made of swarming
filamentous bacteria. We simultaneously measure orientation and
velocity fields and show that the complex spatiotemporal dynamics
of our system can be quantitatively reproduced by a new type of microscopic
model for active suspensions whose important parameters are all estimated from
comprehensive experimental data. This provides unprecedented access
to key effective parameters and mechanisms governing active nematics. 
Our approach is applicable to different types of dense suspensions and
shows a path towards more quantitative active matter research.
\end{abstract}

\dates{This manuscript was compiled on \today}
\doi{\url{www.pnas.org/cgi/doi/10.1073/pnas.XXXXXXXXXX}}

\begin{document}

\maketitle
\thispagestyle{firststyle}
\ifthenelse{\boolean{shortarticle}}{\ifthenelse{\boolean{singlecolumn}}{\abscontentformatted}{\abscontent}}{}

\dropcap{E}xamples of active matter can be found at diverse lengthscales \cite{Ramaswamy2010,vicsekreview,RevModPhysMarchetti,RevModPhys.88.045006,Elgeti2015,Needleman2017}
from animal groups 
\cite{Cavagna2010,Cavagna2014,Buhl2006,Berdahl2013,Ouellette2015} 
to cell colonies and tissues 
\cite{Dombrowski2004,Sokolov2007,Zhang2010a,PhysRevE.95.020601,Chen2017b,Dapeng2016,Yang2017tissue},
to {\it in vitro} cytoskeletal extracts 
\cite{Schaller2010,Sumino2012,Sanchez2012,Keber2014,Guillamat2016a,Wu2017,Suzuki2017b,Guillamat2016a,Ellis2017a}, 
and man-made microscopic objects 
\cite{Palacci2013,Yan2016,Bricard2013,Theurkauff2012,Jiang2010a,Buttinoni2013}. 
Energy input at the level of the individual constituents drives active matter systems
out of thermal equilibrium and leads to a wide range of collective phenomena, 
including flocking \cite{Cavagna2010,Yan2016,Bricard2013,Deseigne2010,Schaller2010,Kumar2014}, 
swarming \cite{Dombrowski2004,Sokolov2007}, 
clustering \cite{Palacci2013,Zhang2010a,Theurkauff2012,Buttinoni2013},
two-dimensional long-range order \cite{PhysRevLett.110.208001,PhysRevE.95.020601}, giant number fluctuations \cite{PhysRevLett.110.208001,PhysRevE.95.020601,Narayan2007,Deseigne2010,Zhang2010a},
spontaneous flow \cite{Lushi2014,Wioland2016,Sanchez2012,Wu2017,Suzuki2017b}, and synchronization \cite{Chen2017b}. 

Active matter systems consisting of elongated particles often lead to local nematic orientational order.
This important {\it active nematics} class comprises experiments with vibrating granular
rods \cite{Narayan2007}, crawling cells \cite{Duclos2016,Saw2017,Kawaguchi2017}, 
swarming sperms \cite{Creppy2015}, filamentous bacteria \cite{PhysRevE.95.020601}
and motor-driven microtubules \cite{Sanchez2012,Sumino2012,Wu2017}, 
which, together with theoretical work, 
have shown that the interplay between orientational order, active stress, particle
and fluid flow leads to complex spatial-temporal dynamics and unusual fluctuations.
The seminal work by the group of Dogic \cite{Sanchez2012,DeCamp2015} has been particularly influential. They experimentally observed spontaneous chaotic dynamics driven by topological defects, and their results triggered
a large number of theoretical and modeling approaches. These are of two main types,
particle-level `microscopic' models \cite{Shi2013,Gao2015,DeCamp2015} 
and continuous-level `hydrodynamic' descriptions 
\cite{Ramaswamy2003,Baskaran2008,PhysRevE.88.050502,Ngo2014,Giomi2013a,Thampi2013,Blow2014,Giomi2015},
with the latter usually written phenomenologically or by complementing equilibrium liquid crystal 
theories with minimal active terms.
These studies provided important insights into the multi-faceted dynamics
of active nematics, such as hydrodynamic instabilities, long-range correlations, anomalous fluctuations, defect dynamics,
spatial and temporal chaos. 
However, these models generally contain a large number of parameters.
This has made comparisons between models and experiments semi-quantitative at best. 

\begin{figure*}[htb]
\includegraphics[width=\textwidth]{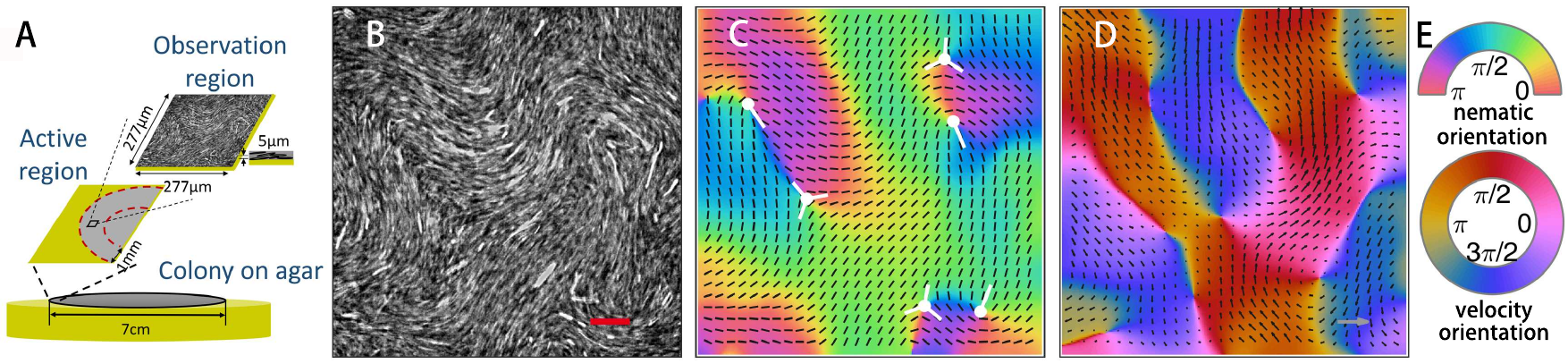}
\caption{(A) Schematic illustration of a bacteria colony growing on agar; 
the active nematics dynamics studied here takes place in a millimeter-wide, few-micron thick region at the edge;
our observation region is typically $277\times 277\mu{\rm m}$.
Mean cell length varies from $6$ to $14 \mu m$, depending on antibiotic concentration.
(B) raw image of our fluorescent cells (scale bar 30$\mu m$) in some experiment performed at 45 $\mu g/mL$ drug level.
(C,D) nematic order and velocity fields extracted from B.
Black rods in C represent unit length director vector $\hat{\bf u}$, and length of arrows in D is proportional to the local speed (scale bar 250 $\mu m/s$).
(E): colormaps coding the orientation of the nematic and velocity fields in C and D.
The white symbols in C represent the $\pm\frac{1}{2}$ defects (see SI text and Fig.~S2 for details about their orientation).}
\label{fig1}
\end{figure*}

Bacteria are widely used as model systems to study active matter
\cite{Dombrowski2004,Sokolov2007,Zhang2010a,PhysRevE.95.020601,Chen2017b}.
A recent study showed that elongated E. coli cells strongly confined 
between two glass plates can display the long-range nematic order and anomalous fluctuations
typical of dry, dilute active nematics systems \cite{PhysRevE.95.020601}. 
However, so far, almost no bacterial system has been reported to exhibit the phenomenology of dense, wet active nematics
as reported first by Dogic {\it et al.}.  One exception is a study of motile bacteria dispersed in a nontoxic lyotropic nematic
liquid crystal \cite{Zhou2014,Genkin2017}. When bacteria concentration
is high enough, active stress destabilizes the ordered nematic state
of this biosynthetic system, leading to a state where topological defects in the liquid crystal evolve chaotically 
in a manner closely resembling that of the Dogic system.
Here, we show that the typical phenomenology of wet, dense, active nematics can
be experimentally realized in colonies of filamentous bacteria 
and show how to build a data-driven quantitatively-faithful theoretical description of it.
To this aim, we introduce a new type of microscopic model for active suspensions and we use 
simultaneous experimental measurements of both orientation and velocity fields to estimate all its parameters.

\section*{Experimental results}

Our experiments are carried out with \textit{Serratia marcescens} bacteria. 
At the edge of growing colonies, 2-3 layers of cells
actively swim by rotating flagella in a micrometer-thick millimeters-wide film of liquid on the agar surface (Fig.~1A). 
Apart from a narrow ($\sim 100\mu$m) outer ring, the thickness of this quasi-2D suspension is very
constant. No obvious spatial or temporal inhomogeneity is noticeable in measured fields.
A sub-lethal level of antibiotic drug Cephalexin is added into the growth agar medium. The drug allows
bacteria to grow but not to divide, leading to long cells. By
varying the drug concentration, we can change the mean cell length by a factor
of two, see Fig.~S1A. 
Bacteria are labeled with a green fluorescent protein, which allows to record their motion under the microscope. 
In the dense, thin layer of interest, cells are almost always in close contact and nearly cover the whole surface. 
Our elongated cells are also frequently nematically aligned, as testified by the presence of $\pm\frac{1}{2}$ 
charge topological defects typical of 2D nematics (Fig.~1B). 
(Standard cells cultivated without antibiotic drug do not give rise to any significant local order.)
Our images do {\it not} allow to distinguish the current polarity of each cell, i.e. in which direction it is currently
swimming with respect to the fluid. In fact, the swimming of most bacteria is strongly hampered at such high density.
Nevertheless, our cells move collectively, mainly advected by the fluid they have set in motion, 
in a spatiotemporally chaotic manner strongly reminiscent of other active nematics systems \cite{Sanchez2012,Zhou2014} (\textcolor{blue}{Movies~S1} and \textcolor{blue}{S2}).
From each image, we extract a nematic orientation field $\hat{\bf u}({\bf r},t)$ through a gradient-based method, 
and ${\bf v}({\bf r},t)$ the velocity field of cells in the laboratory frame using a standard particle image velocimetry technique
(Fig.~1C,D, {\it SI Appendix, Fig.~S2}). Movie~S1 shows the typical evolution of the obtained coarse-grained orientation and velocity fields. 
This dynamics is fast. Typical correlation times are of the order of seconds (see below). 
In each experiment, we record images for 30 s, which is significantly shorter than the cell division time (20 min). Therefore, contributions of cell growth to active stress are negligible in our work \cite{Doostmohammadi2016b,C5SM01382H}.

\begin{figure*}[htb]
\includegraphics[width=\textwidth]{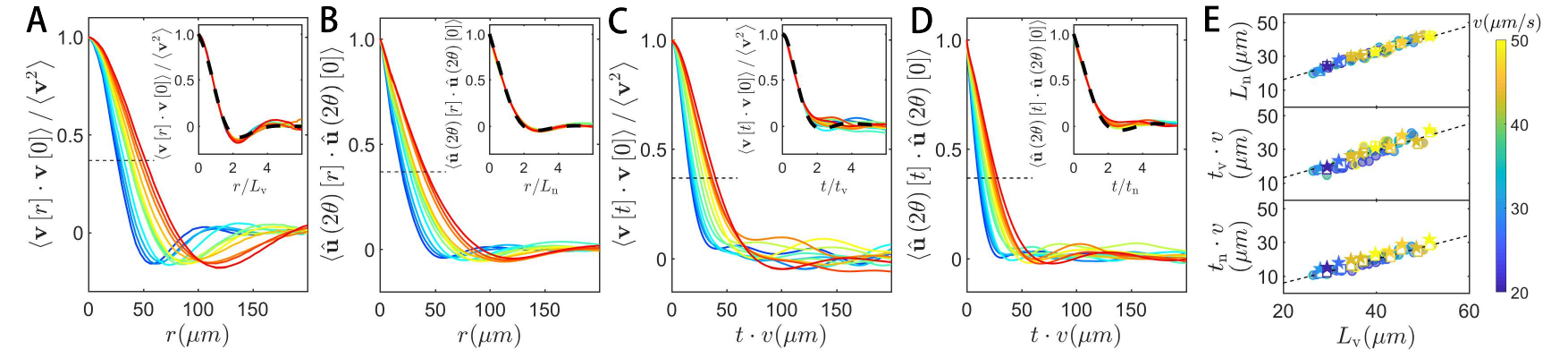}
\caption{Spatial (A and B) and temporal (C and D) two-point correlation functions (defined in y-axis labels) from
experiments with drug concentrations from 30$\mu g/mL$ to 60$\mu g/mL$. 
The color of each curve codes for the velocity correlation length $L_{\rm v}$.
Insets: correlation functions rescaled by their correlation length/time.
Dashed lines: correlation functions calculated from numerical simulations of the model performed at the optimal-match parameter values
estimated for the typical 45 $\mu g/mL$ experiment shown in Fig.~1, corresponding to 6th column in {\it SI Appendix, Table~S1}.
(E): variation of various correlation lengths with the velocity correlation length $L_{\rm v}$ for both experiments (solid circles and open squares)
and simulations of our model at optimal parameter values (solid stars). 
The experiments noted by open squares are those corresponding to the optimal model simulations noted with solid stars.
Each symbol is colored according to the mean speed $v$ measured (colormap at the right).}
\label{fig2}
\end{figure*}

\subsection*{Global measurements}

We first measure global statistical properties of our velocity
and orientation fields.
The average cell speed  $v\equiv\langle|{\bf v}({\bf r},t)|\rangle_{{\bf r},t}$ 
varies between 20 and $50\text{ }\mu m/s$ from experiment to experiment 
but is approximately independent of the drug concentration ({\it SI Appendix, Fig.~S1B,C}).

Next we compute spatial and temporal two-point correlation functions, which are defined and shown in Fig.~2A-D. 
The spatial/temporal separations
corresponding to a correlation value of $1/e$ are identified as the correlation lengths and times. 
Symbols $L_{\rm v}$, $L_{\rm n}$, $t_{\rm v}$, and $t_{\rm n}$ respectively denote velocity and orientation correlation lengths,
velocity and orientation correlation times. 
These quantities are typically of the order of tens of $\mu m$ and one second.
When we increase the cell length with antibiotics, the correlation lengths $L_{\rm n}$ and $L_{\rm v}$ increase systematically (Fig.~2A,B).
Such a systematic variation is only observed for correlation times $t_{\rm n}$ and $t_{\rm v}$ 
if time is rescaled by the mean speed $v$ (Fig.~2C,D).
Correlation functions from various experiments with different drug concentrations 
collapse onto each other when space and time are rescaled by correlation lengths and times (insets in Fig.~2A-D).
Moreover, all these quantities are linearly related to each other. 
Strikingly, transforming correlation times into correlation lengths using the mean speed $v$,
we find that $L_{\rm n}$, $v \cdot t_{\rm n}$, and $v \cdot t_{\rm v}$  are all proportional to $L_{\rm v}$ with approximately
the same slope (Fig.~2E). 
This indicates that our experiments are characterized by a single lengthscale and the mean flow speed
 \cite{Zhang2009,sokolov2012}.
Because our bacteria are too closely packed to measure their length,
we use $L_{\rm v}$ and $v$ as ``effective control parameters'' of our experiments, with 
$L_{\rm v}$ serving as a good proxy to the mean cell length, see {\it SI Appendix, Fig.~S1}.

\begin{figure}[t]
\includegraphics[width=\columnwidth]{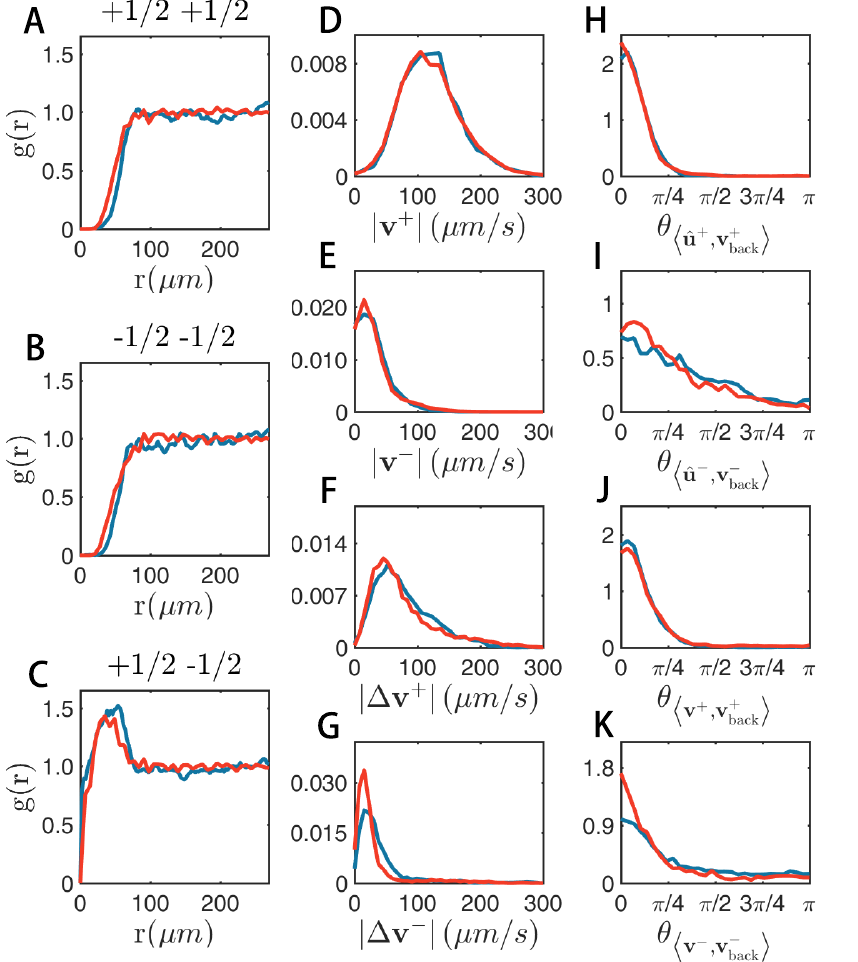}
\caption{Statistical properties of defect cores. Experimental data extracted from the experiment
at drug level 45$\mu g/mL$ 
that gives a correlation length $L_{\rm v} = 39.7\mu m$ and a mean flow speed $v=50.4\mu m/s$ 
used in most of the text (blue curves), and from simulations of our model at parameter
values optimized for that experiment (red curves), corresponding to 6th column of {\it SI Appendix, Table~S1}.
(A-C): Two-point pair correlation functions $g(r)$ for the positions of defect cores (respectively 
$(+\frac{1}{2},+\frac{1}{2})$, $(-\frac{1}{2},-\frac{1}{2})$, and $(+\frac{1}{2},-\frac{1}{2})$ pairs).
(D-G): probability distribution functions of various defect speeds 
(respectively speed of $+\frac{1}{2}$ and $-\frac{1}{2}$ defects in lab frame, and 
speed of $+\frac{1}{2}$ and $-\frac{1}{2}$ defects in fluid frame).
(H,I): probability distribution functions of angle between defect orientations $\hat{\bf u}^\pm$ and fluid velocity at their core ${\bf v}^\pm_{\rm back}$.
(J,K): same as (H,I), but for the defect core velocities in the lab frame ${\bf v}^\pm$.}
\label{fig3}
\end{figure}

\subsection*{Defect properties}

To go beyond the reduction of the complex spatiotemporal dynamics of our bacterial system to just a lengthscale and the mean speed,
we now focus on the $\pm\frac{1}{2}$ topological defects of the orientation field. 
Their detailed structure and their dynamics offer unique access to the coupling between nematic order and flow, 
all information that we will show later to be crucial to  determine model parameters.

We identify the location of $\pm\frac{1}{2}$ defects by contour integral of the director field (see Fig.~1C and Movie~S1 for typical results). 
From the trajectories of defect cores, we measure ${\bf v}^\pm$, their velocity in the lab frame.
We also measured the velocity of defects in the fluid frame, 
$\Delta {\bf v}^\pm={\bf v}^\pm - {\bf v}^\pm_{\rm back}$, where ${\bf v}^\pm_{\rm back}$ is the fluid ``backflow'' velocity 
averaged over a small region surrounding the defect core (see {\it SI Appendix, Fig.~S2D}).
We finally determine the intrinsic orientation $\hat{\bf u}^\pm$ of defects. This is straightforward for the comet-shaped $+\frac{1}{2}$ defects. 
For the $-\frac{1}{2}$ defects, which are not polar but have a three-fold symmetry with three radial axes along which the nematic director is aligned, 
we choose the axis closest to the current orientation of ${\bf v}^-$ (see {\it SI Appendix} and {\it SI Appendix, Fig.~S2} for details).

As in other active nematics systems \cite{Narayan2007,Sanchez2012,Kawaguchi2017}, defects are created in $\pm$ pairs 
via the bending of ordered regions (\textcolor{blue}{Movie~S1}).
Upon generation, $+\frac{1}{2}$ defects typically quickly move away and
less motile $-\frac{1}{2}$ defects stay longer near the generation site. 
Pair of defects of opposite charge may also annihilate upon encounter.
In a given experiment, generation and annihilation of defects balance
each other so that their total number is approximately constant in time.
The radial distribution functions of defect position, $g\left(r\right)$, reveals that 
defects with the same sign repel from each other at short distances (Fig.~3A-C). 
For defects of opposite sign, $g(r)$ has a short-scale peak reflecting the fact that defects are created in $\pm$ pairs \cite{DeCamp2015}.

Restricting our analysis to ``isolated'' defects from now on, 
i.e. whose distance from nearest neighbors is larger than nematic correlation length $L_{\rm n}$,
we observe that they are essentially distributed randomly in space: no global translational nor orientational
order is observed. 
Defect speed distributions, both in the lab and in the fluid frame, show that $+\frac{1}{2}$ defects are more motile, but 
$-\frac{1}{2}$ defects do {\it not} have a negligible speed, even in the fluid frame (Fig.~3E,G).
We also find that the defect orientation $\hat{\bf u}^\pm$ is strongly correlated to their velocity orientation, 
and to the orientation of their velocity in the fluid frame. 
Essentially, all three vectors are aligned, even for the $-\frac{1}{2}$ defects (Fig.~3H-K). 
Note that a small but finite velocity in the fluid frame $\Delta {\bf v}^-$ is at odds with usual statements about $-\frac{1}{2}$ defects in active nematics, 
where they are treated as symmetric, force-free, diffusive objects \cite{Giomi2014,Giomi2015,Shi2013}. We elaborate on this point in Discussion section. 

To further quantify the structure of defects, we average, over time and many defects, the orientation and
velocity fields around their core, sitting in their intrinsic reference frame. The familiar mushroom-shape and three-fold symmetry
of, respectively, the $+\frac{1}{2}$ and $-\frac{1}{2}$ defect are clearly observed (Fig.~4A,B). 
The flow field around the $+\frac{1}{2}$ defect core shows a strong jet, while 3 nearly symmetric jets go through the center
of the $-\frac{1}{2}$ defect (Fig.~4C,D), in agreement with previous work \cite{Giomi2014,Giomi2015}. 

Because of the chaotic collective dynamics, the magnitude of these averaged fields decays away from the defect core.
We define defect core sizes $R^{\pm}$ as the radius where the magnitude of averaged director vector $\left|\textrm{u}_{\rm a}\right|$ reaches value $\frac{1}{2}$.
For the quantitative modeling of our system, we also 
extracted angular profiles of orientation and velocity around defect cores from the averaged fields.
In Fig.~4E,F, we plot profiles of the angle of the nematic director
calculated at three different radii around the defect cores.
These profiles show clear systematic deviations from
the linear variation predicted in one-constant equilibrium liquid crystals theory \cite{ZhangRui2017}. 
The velocity orientation profiles, as well as the profiles of the magnitude of orientation and velocity fields
show also systematic variations reflecting the fine-structure of defects (Fig.~4I-L). 

\begin{figure*}[t]
\includegraphics[width=\textwidth]{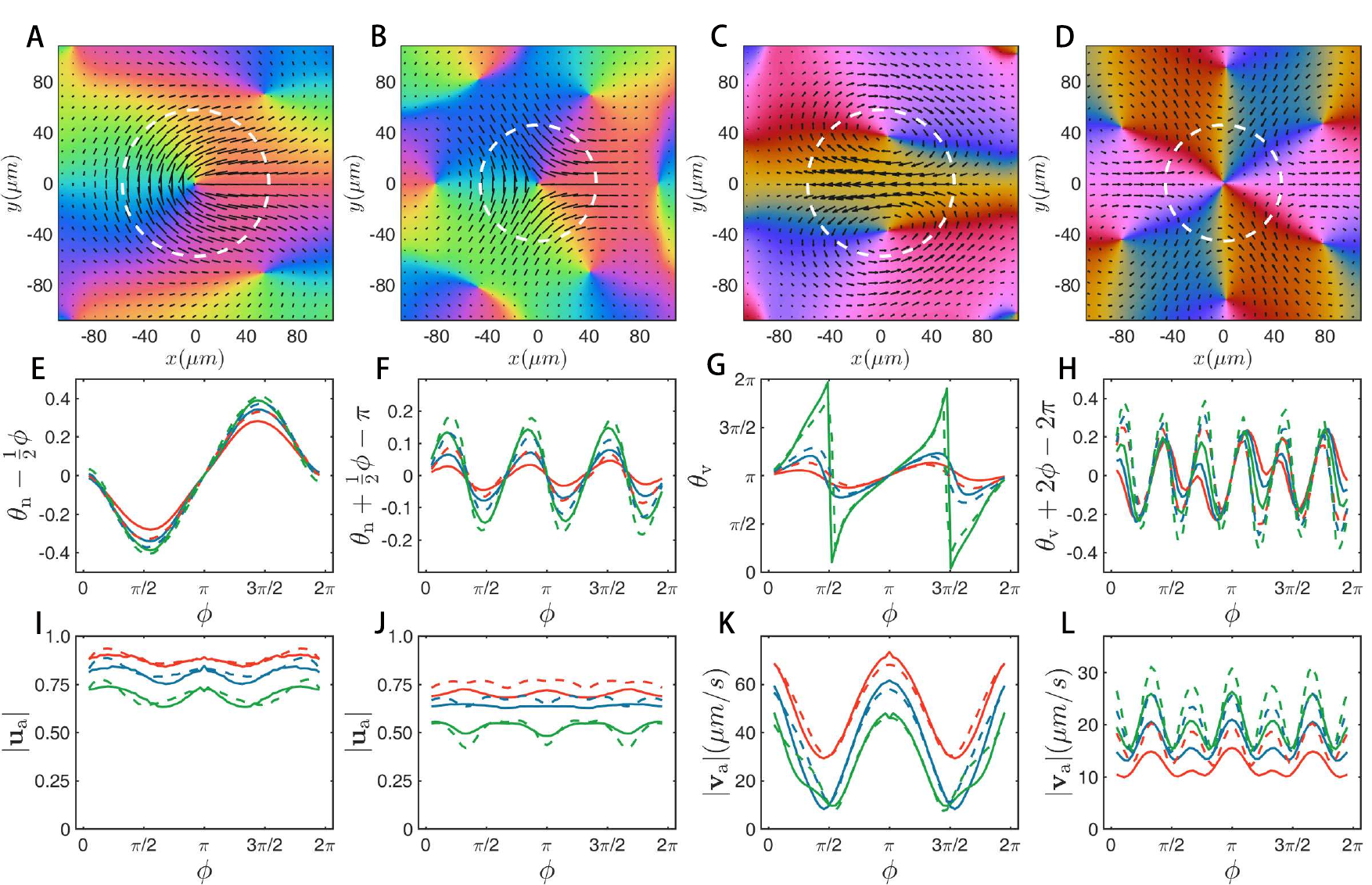}
\caption{Mean nematic (columns 1 \& 2) and velocity (columns 3 \& 4) fields around isolated 
$+\frac{1}{2}$ (columns 1 \& 3) and $-\frac{1}{2}$ (columns 2 \& 4) defects. 
(A-D): full 2D representation as in Fig.~1. The white circles show the defect sizes $R^+$ and $R^-$ defined in the text.
Panels in the second (E-H) and third (I-L) rows contain angular profiles of the
2D fields measured at three different radii: $23\mu m$ (red), $33\mu m$ (blue), $43\mu m$ (green). 
Angle $\phi$ is the angle depicting the circles of various radii around the defect cores. 
Mean director $\theta_{\rm n}$ and velocity $\theta_{\rm v}$ angles are defined as the angles between mean director/velocity vectors and x axis.
The reference linear component ($\frac{1}{2}\phi$ in E, $\pi-\frac{1}{2}\phi$ in F, $2\pi-2\phi$ in H) has been 
substracted in the orientation profiles to emphasize deviations from single-Frank constant liquid crystal theory.
Solid lines: angular profiles from experiments. Dashed lines in E-L: angular profiles from simulations performed at optimal parameters. 
}
\label{fig4}
\end{figure*}

\begin{figure*}[t]
\includegraphics[width=\textwidth]{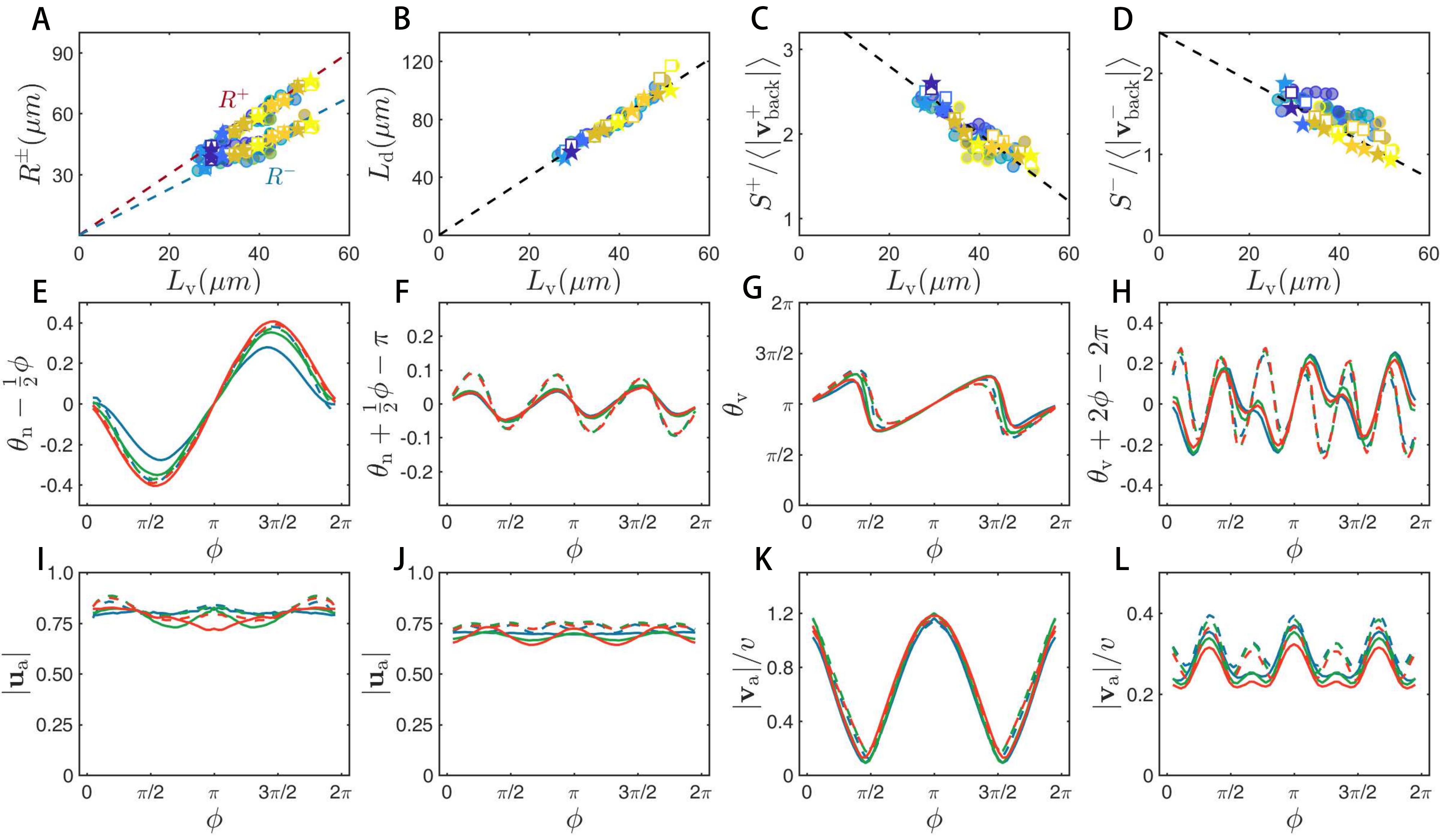}
\caption{Dependence of defects properties on experimental conditions. 
(A-D): variation with nematic correlation length $L_{\rm v}$ of defect core size $R^\pm$ (A), mean inter-defect lengthscale $L_{\rm d}$ (B), 
mean of $+\frac{1}{2}$ and $-\frac{1}{2}$ defect speed $S^{\pm}$ divided by local flow speed $\left<|{\bf v}^\pm_{\rm back}|\right>$ (C,D).
Experiment and simulation results are shown by open squares and
solid stars, color-coded by the mean speed $v$, following Fig.~2E. All experiments and simulations collapse onto single straight lines, indicating that $v$ does not influence the quantities represented.
(E-L) angular profiles as in Fig.~4 calculated along the circle of radius 0.6$R^\pm$ 
for 3 experimental datasets obtained at different drug levels 
(blue: 30$\mu g/mL$, 1st column in {\it SI Appendix,Table~S1}, green: 45$\mu g/mL$, 6th column in {\it SI Appendix, Table~S1}, red: 60$\mu g/mL$,12th column in {\it SI Appendix, Table~S1}). Corresponding simulation results are shown using dashed lines.} 
\label{fig5}
\end{figure*}

We have performed the above analysis of the dynamics and fine structure of defects on a large set of experiments.
We now describe how the main defect properties vary with our two effective control parameters, the correlation length $L_{\rm v}$ 
and the mean flow speed $v$.
The defect core sizes $R^\pm$ vary linearly with $L_{\rm v}$, and are roughly independent of $v$ (Fig.~5A).
In the steady-state, the density of defects is statistically constant. From this steady density one can extract an inter-defect lengthscale $L_{\rm d}$, 
which behaves like all other correlation lengths, in agreement with previous work on wet active nematics \cite{C6SM00812G,Guillamat2017} (Fig.~5B).
We also find that the speed of defects relative to the local flow speed at their core decreases with $L_{\rm v}$ while being also roughly 
independent of $v$ (Fig.~5C,D, see a discussion of this below).
Remarkably, the detailed spatial structure of defects does not vary significantly between experiments with different characteristic lengths: 
after rescaling spatial coordinates by defect core size, or, equivalently, correlation length, averaged director and velocity fields
from different datasets overlap nicely.
We further confirm this by comparing defect angular profiles at 0.6$R_{\pm}$ for different experiments (Figs.~5E-L).

\section*{Quantitative modeling}

A microscopically faithful model of our dense, thin bacterial system where cells and their many flagella are in constant contact with each other and 
with the gel substrate is a formidable task well-beyond current numerical power. 
Besides, this would require the knowledge of many specific details that are unknown. 
Here we adopt a radically different approach: we treat the collisions and local interactions between cell bodies and their flagella 
at some effective level, where, we assume, they amount to a combination of steric repulsion and alignment. 
In addition, the far-field interactions and other effects due to the incompressible fluid surrounding bacteria are taken into account
by solving the Stokes equation for the fluid flow.
All this also allows us to build an efficient, streamlined, but comprehensive model in two space dimensions.

\subsection*{Description of the numerical model}

Recall that most cells in our dense system are not able to swim freely, simply because nearby cells prevent them to do so
(see \textcolor{blue}{Movie S2}).
These crowded cells mostly exert force dipoles on the fluid, which is then set in motion by their collective action. 
Cells, in turn, are advected and rotated by the fluid. 
Our model thus consists of non-swimming force dipoles immersed in an incompressible fluid film, and 
differs significantly from the common choice of using a dynamic equation for a director field \cite{Doostmohammadi2016,Doostmohammadi2018,Thampi2014c,Putzig2016,Srivastava2016a,Guillamat2016}.
As shown by a schematic diagram in {\it SI Appendix, Fig.~S3}, each dipole represents the local cell body orientation and active forcing.

The fluid flow ${\bf v}({\bf r},t)$ is the solution of the (2D) Stokes equation
\begin{equation}
\mu\nabla^{2}{\bf v}+\nabla p-\alpha{\bf v}+{\bf F}=0 \;\;{\rm with}\;\; \nabla\cdot{\bf v}=0
\label{eq1}
\end{equation}
where $\mu$ is the fluid viscosity, $\alpha$ is the effective friction with the substrate, 
$p$ is the pressure enforcing the incompressibility condition, and
${\bf F}$ is the active force field exerted by dipoles on the fluid \cite{Doostmohammadi2016}.

Our dipoles are point particles with position $\mathbf{r}_{i}$ and orientation $\theta_{i}$ (or, equivalently, unit orientation vector 
$\hat{{\bf u}}_{i}=(\cos(\theta_{i}),\sin(\theta_{i}))$.
They locally align, are advected and rotated by the flow, and experience pairwise repulsion to keep their density homogeneous:
\begin{eqnarray}
\dot{{\mathbf r}}_{i} &=& {\bf v}\left({\mathbf r}_{i}\right) + C_{\rm r} \sum_{j\sim i} {\bf R}_{ij} \label{eq3} \\
\dot{\theta}_{i}&=&C_{\rm a} \sum_{j\sim i}\sin\left[2\left(\theta_{j}-\theta_{i}\right)\right] +C_{\rm v}\left(\nabla\times{\bf v}\right)\cdot\hat{{\bf z}} \nonumber \\
&& +C_{\rm s}\hat{{\bf u}}_{i}\times\left({\bf E}\cdot\hat{{\bf u}}_{i}\right)\cdot\hat{{\bf z}}+ C_{\rm n} \xi_{\theta}, \label{eq4}
\end{eqnarray}
In \eqref{eq4}, the first term on the right-hand side, with strength $C_{\rm a}$,  
codes for the nematic alignment of dipole $i$ with all neighbors currently present within distance $R_{\rm a}$.
The next two terms govern how dipoles are rotated by the flow field ${\bf v}$, following Jeffery's classic work: 
both local vorticity $\nabla\times{\bf v}$ and local strain ${\bf E}=\left(\nabla{\bf v}+\nabla{\bf v}^{T}\right)/2$ are playing
a role, but with coefficients $C_{\rm v}$ and $C_{\rm s}$ taking values a priori different from the classic ones
calculated for perfect ellipsoids with no-slip boundaries \cite{Jeffery1922}.
Finally, $\xi_\theta$ is a unit-variance, white, angular Gaussian noise. 
In \eqref{eq3}, the right-hand side term ${\bf R}_{ij}$ represents a unit-range pairwise soft repulsion force between dipoles of strength $C_{\rm r}$.
Note that self-propulsion is not included in \eqref{eq3} because our system is crowded.

The force field ${\bf F}$ in \eqref{eq1} is assumed to be dominated by the gradient of the active stress tensor field. (A small, residual contribution from the short-range repulsion force between neighboring dipoles exists, but can usually be neglected, see {\it SI Appendix, Eq.~S2} for details about this point.) The active stress tensor is itself assumed, as usual in wet active nematics studies \cite{Giomi2013a,Thampi2013}, to be proportional to the gradient of 
the orientation field:
\begin{equation}
{\bf F}=f_0 \nabla \cdot \hat{{\bf u}}\hat{{\bf u}}
\label{eq5}
\end{equation}
where $f_0$ is the typical strength of dipoles. In experiments, $\hat{{\bf u}}$ is the {\it measured} nematic orientation field. 
In the model, $\hat{{\bf u}}$ is the local coarse-grained orientation of our dipoles.

\begin{figure}[htb]
\includegraphics[width=\columnwidth]{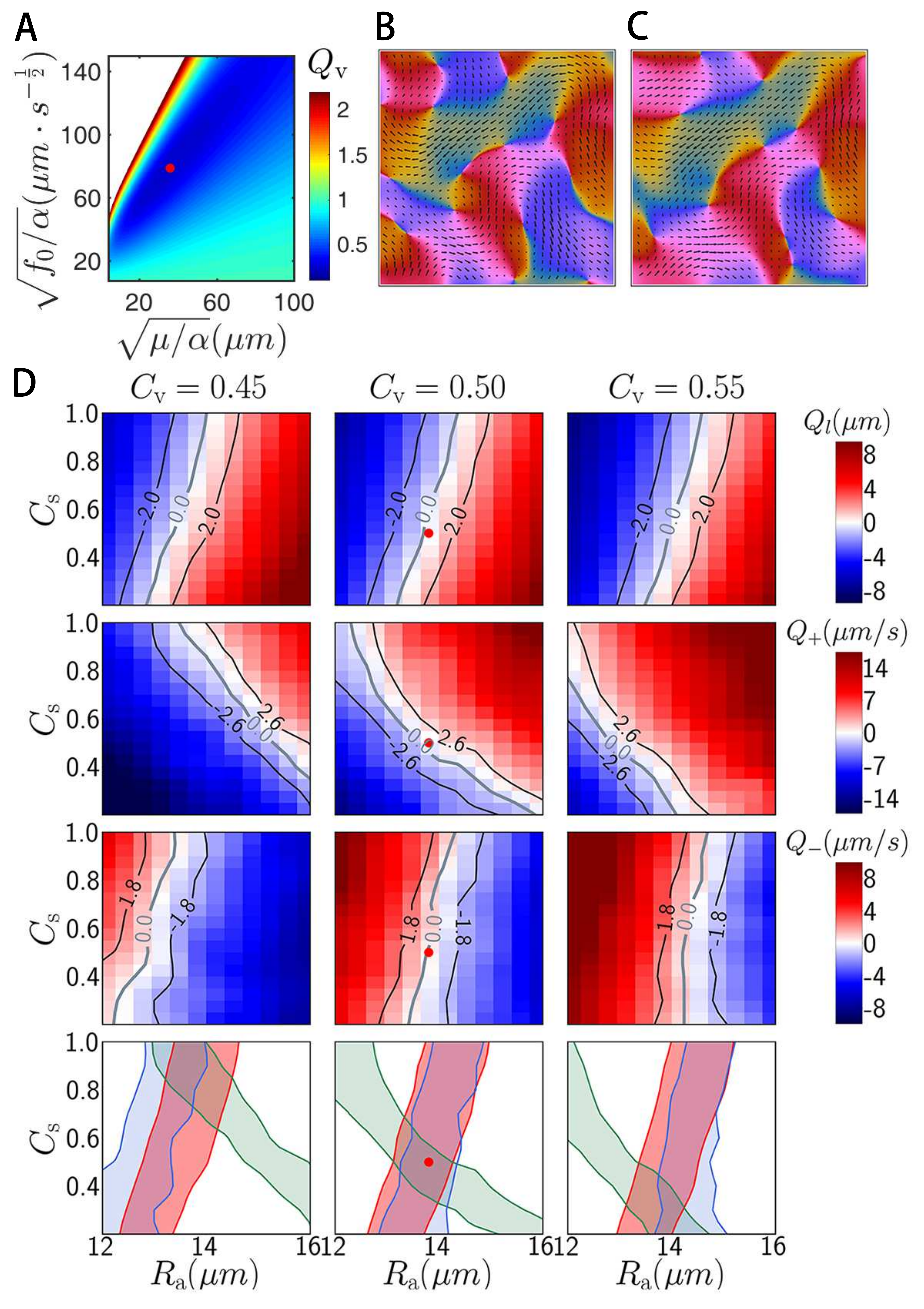}
\caption{Quality functions and results from hydrodynamic (A-C) and full (D)
model matching. Optimal parameters are marked by a red dot. 
(A): quality function $Q_{\rm v}$ in the $(\sqrt{\mu/\alpha},\sqrt{f_0/\alpha})$ plane.
(B): a typical instantaneous experimental velocity field.
(C): velocity field reconstructed from the orientation field measured at the same time as (B) at the optimal parameters
indicated in (A).
The first three rows of (D) contain quality functions of
characteristic length ($Q_{\rm l}$), +1/2 defect speed ($Q_{+}$) and
-1/2 defect speed ($Q_{-}$). Each panel represent a scan in the $C_{\rm s}$-$R_{\rm a}$
plane and panels in the same column use the same $C_{V}$ value. Black
contour lines mark regions of acceptable deviations. Panels
in the last row contain acceptable regions of parameters extracted
from quality functions: red from $Q_{\rm l}$, green from $Q_{+}$ and
blue from $Q_{-}$; they show that $C_{\rm v}=0.5$ yields largest overlap
area for acceptable regions of parameters and a red dot in the middle
panels marks the optimal choice for parameters $C_{\rm s}$ and $R_{\rm a}$. }
\label{fig6}
\end{figure}

The full system constituted by Eqs.~[1-4] can be seen as a minimal Vicsek-style model \cite{VICSEK1995,Chate2006} incorporating the main mechanisms at play in our bacterial active nematics.
One thus expects a basic interplay between alignment and noise: if the alignment strength
$C_{\rm a}$, or the alignment range $R_{\rm a}$, or the number density of dipoles $\rho_0$ is large enough, 
or if the noise strength $C_{\rm n}$ is weak enough, local orientational (nematic) order can emerge.
The global number density of dipoles $\rho_0$ and the noise strength $C_{\rm n}$ have opposite effects.
We checked that changing $\rho_0$ in the experimentally reasonable range $[1.5,4]$ (in simulation units) 
yields similar results. In the following, we fix $\rho_0=1.5$ to lighten the numerical task.

It is  relatively easy to find parameter values such that the dynamics of 
our model closely resembles the experimental observations.  As a matter of fact, the region of parameter space where 
spatiotemporally chaotic active nematics behavior occurs is rather large. To go beyond such qualitative agreement, we have systematically investigated
the effects of parameters. We now show that for each experimental dataset, there exists a unique set of parameter values
at which the model optimally matches the experiment, in the sense that all quantities studied in the previous section are in quantitative agreement. 

\subsection*{Data-driven parameter optimization}

We proceed in two steps. 
First, simultaneous measurements of velocity and orientation fields allow us to pinpoint the parameters in  \eqref{eq1} without
resorting to the "microscopic" part of the model, i.e. Eqs.~[2,3].

Dividing both sides of \eqref{eq1} by $\alpha$, we are left with two independent parameters,
$\mu/\alpha$ and $f_{0}/\alpha$. 
Therefore, for any given pair of parameters $\mu/\alpha$ and $f_{0}/\alpha$, 
and a particular experimentally-measured orientation field $\hat{{\bf u}}$,
we can compute the velocity field ${\bf v}_{*}$ solution of \eqref{eq1}.
We then compare ${\bf v}_{*}$ with ${\bf v}$, the velocity field measured at the same time as $\hat{{\bf u}}$. 
Scanning the whole $(\mu/\alpha, f_{0}/\alpha)$ parameter plane, we find that there is an optimal point where
the difference between ${\bf v}_{*}$ and ${\bf v}$ is minimal on average. Specifically, we measure the quality function
$
Q_{\rm v}(\mu/\alpha, f_{0}/\alpha) = 
\langle |{\bf v}_{*}({\bf r},t)-{\bf v}({\bf r},t)|^{2} / |{\bf v}({\bf r},t)|^{2} \rangle _{{\bf r},t},
$
where the average is carried out over both space and time. A typical result
for an experiment with 45$\mu g/mL$ drug concentration is in Fig.~6A, where $Q_{\rm v}$ shows a minimum for
$Q_{\rm v}=0.23$ at $f_{0}/\alpha=6174\mu m^{2}/s$ and $\sqrt{\mu/\alpha}=36\mu m$. Typical instantaneous velocity fields ${\bf v}_{*}$ produced at these parameters compare very well to the corresponding ${\bf v}$ fields
(Fig.~6B,C, \textcolor{blue}{movie~S3}).

After fluid parameters are fixed, we proceed to the second step and match the full model with experiments. 
Eqs. (2-3) contain six parameters. We first evaluate their influence by varying them individually around a reference point 
(see {\it SI Appendix}, and {\it SI Appendix, Fig.~S4} for details). We find that angular noise level $C_{\rm n}$ and repulsion strength $C_{\rm r}$ are not sensitive parameters provided local order is not destroyed by strong noise and particles do not crystallize for too-strong repulsion. We therefore fix $C_{\rm n}=1.0$ and $C_{\rm r}=0.5$. Nematic alignment parameters $C_{\rm a}$ and $R_{\rm a}$ play a major, 
but similar role, so we decide to fix  $C_{\rm a}=0.4{\rm s}^{-1}$ and vary $R_{\rm a}$, mimicking the change of cell length in experiments.
This leaves us with only three parameters to vary, $R_{\rm a}$, $C_{\rm v}$, and $C_{\rm s}$,
when looking for an optimal match between model and experiment.

We performed a systematic 
scan of this restricted parameter space, running the model for many sets of parameter values, and extracting from each of these runs the quantities
of interest, i.e. those measured also in the experiment. 
To quantify the match between model and experiment, we found that using 3 independent quality functions is sufficient. 
Here we use $Q_{\rm l} \equiv L_{\rm n*}-L_{\rm n}$, the difference in nematic correlation length,
$Q_{+} \equiv S_{*}^{+}-S^{+}$ and $Q_{-} \equiv S^{-}_{*}-S^{-}$ the differences in defect speed ($S^{\pm} \equiv \langle|{\bf v}^{\pm}|\rangle$), respectively for the $+\frac{1}{2}$ and $-\frac{1}{2}$ defects. (As before, the $*$ subscript denotes quantities measured on the model.)
Computed quality functions in the three-dimensional parameter space \{$R_{\rm a}$, $C_{\rm v}$, $C_{\rm s}$\}
are shown in Fig.~6D. 
Perfect matching ($Q_{\{\rm l,+,-\}}=0$) occurs for each function on a surface. 
These three surfaces approximately cross at a single point, as shown in the last row of Fig.~6D.
For the particular experiment considered, we find $R_{\rm a}=13.8\mu m$, $C_{\rm v}=0.5$, and $C_{\rm s}=0.5$,
which thus defines our optimal set of model parameters.
By construction, these parameter values optimize the match between model and experiment for what concerns the 
quantities involved in the quality functions used. Remarkably, we observe that all other quantities not used in these functions are also
quantitatively matched. This is in particular the case for all correlation functions in Fig.~2, all distributions
of defect speed and orientation, and spatial distributions of defects in Fig.~3, all averaged angular profiles
of isolated defects in Fig.~4, defect size and speed in Fig.~5 (see also simulations of the model at optimal parameters in \textcolor{blue}{Movie~S4}).

Finally, we performed two "consistency checks". We verified that choosing a different value of $C_{\rm a}$ yields a different optimal value of $R_{\rm a}$ 
but that all other optimal parameter values then approximately remain the same ({\it SI Appendix, Fig.~S6}). In short, $R_{\rm a}$ and $C_{\rm a}$ are fully redundant.
Next, taking our optimal parameter set, but now ``freeing'' the fluid parameters $\mu/\alpha$ and $f_{0}/\alpha$
from the values determined during our first step, 
we find that these initial values remain optimal ({\it SI Appendix, Fig.~S7}).
This confirms that our procedure, for a given experiment, yields a unique set of model parameters
at which model dynamics optimally matches spatiotemporal data.

\subsection*{Variation of model parameters}

We have successfully applied our matching procedure to a large set of experiments with drug concentration above 15 $\mu g/mL$, 
the level below which cells are too short to give rise to a clear local nematic orientation that can be reliably measured.
For each experiment, the quality of the matching between experiments and simulations
remains excellent. 
Corresponding orientation and velocity fields from simulations at these optimal parameters are shown in Movie~S4.
We thus obtained the variation of the optimal model parameter values with the two experimental effective control parameters, 
the correlation length (proxy for cell size) and the mean flow speed $v$ (see Fig.~7 and {\it SI Appendix, Table~S1}). 
This provides us with a wealth of information about our experimental system.
 
We first discuss the effect of the mean speed $v$ at fixed correlation length. Choosing a subset of experiments
yielding approximately the same correlation length, we observe that $v$ almost exclusively influences $f_0$, 
and does so linearly (Fig.~7A).
The other parameters remain constant with the exception of the interaction range $R_{\rm a}$, which grows slightly with $v$ (Fig.~7B-E).
The clear linear growth of $f_0$ confirms that, via $v$, one has direct access to the the strength of forces dipoles, 
which, in turn, can be interpreted to be proportional to the power developed on average by each flagellum.
As for the weaker linear variation of $R_{\rm a}$ with $v$, we attribute it to the fact that for higher $v$, which corresponds to higher $f_0$, the fluid flow would be destabilized faster, leading to a smaller correlation length. Increasing $R_{\rm a}$ compensates for this.

The variation of optimal model parameters with correlation length, at fixed mean speed $v$, is presented in Fig.~7F-J.
From the extracted ``fluid'' parameters, we can construct two length
scales that are proportional to $L_{\rm v}$. Balancing the active force
term $\nabla\cdot\left(f_{0}\hat{{\bf u}}\hat{{\bf u}}\right)$ in
\eqref{eq1} with the friction term $\alpha {\bf v}$, we have $f_{0}/(\alpha\left|{\bf v}\right|)\sim L_{\rm n}$
which leads to $f_{0}/(\alpha\left|{\bf v}\right|)\sim L_{\rm v}$.
This is confirmed in Fig.~7F. We can also balance friction with
the viscous force $\mu\nabla^{2}{\bf v}$ and get $\sqrt{\mu/\alpha}\sim L_{\rm n}$, as shown in Fig.~7G. 
(We show that the two scalings above are verified for all our data points in {\it SI Appendix, Fig.~S8}.) 
These findings provide a physical understanding of the factors contributing to the correlation length, 
and in particular of how it is connected to the fluid effective parameters.
The vorticity coupling parameter $C_{\rm v}$ is approximately a constant, $C_{\rm v}\simeq0.5$ (Fig.~7H). 
This is in agreement with Jeffery's theory, which shows that $C_{\rm v}=\frac{1}{2}$ 
for almost any axisymmetric shape from needles to ellipsoids to disks. Nearly constant $C_{\rm v}$ is also consistent with observations that defect shape changes little in different experiments (Fig.~5E-L) and that $C_{\rm v}$ is closely connected to defect shape in our model ({\it SI Appendix, Fig.~S5}).
On the other hand, the strain coupling parameter $C_{\rm s}$ decreases with $L_{\rm v}$ (Fig.~7I), at odds with Jeffery's results, 
which show that longer objects have higher $C_{\rm s}$.
This can be understood by noticing that in the model particles do not represent cells. 
Rather, over the interaction range $R_{\rm a}$, several dipoles stand for a cell. They react individually to
the local strain, and thus their response must be weaker than that of a cell, and the longer the cell, the weaker the response. Finally, the range of nematic alignment $R_{\rm a}\sim\sqrt{L_{\rm v}}$ (Fig.~7J), which shows that the correlation length increases linearly with the area where nematic alignment takes place, i.e the number of aligning neighbors, in our model.

\begin{figure}[tb]
\includegraphics[width=\columnwidth]{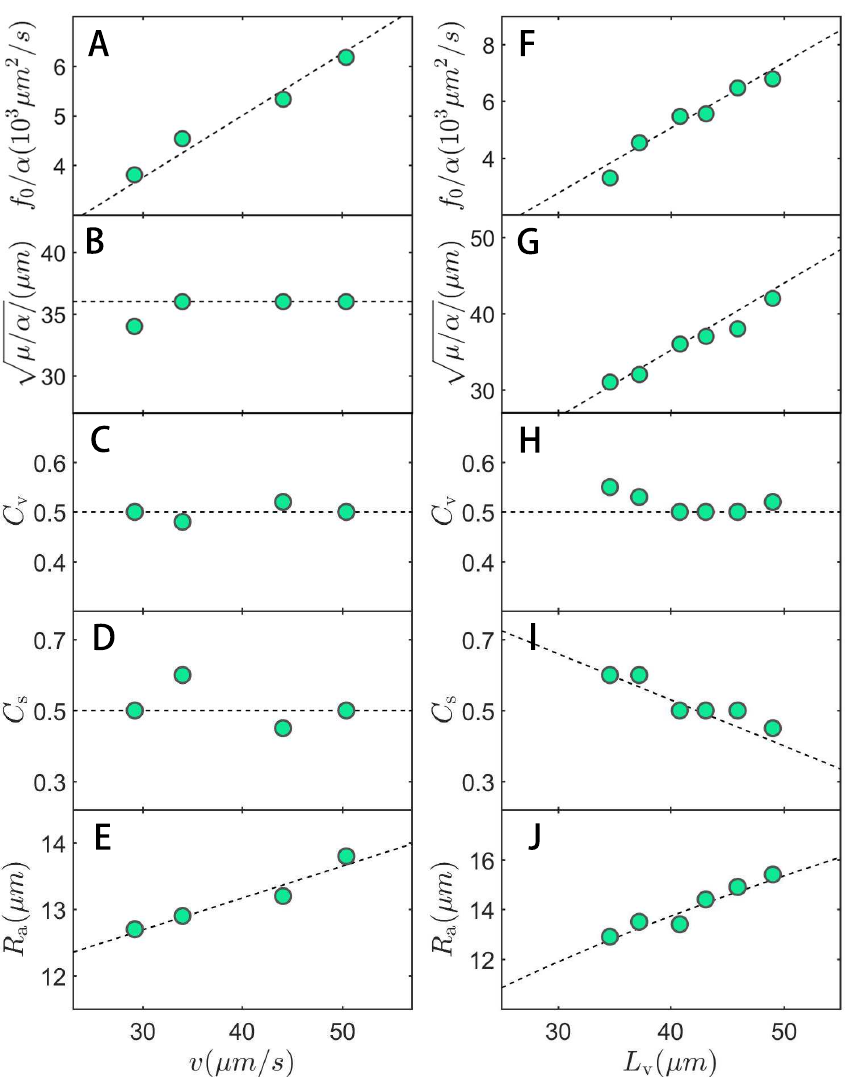}
\caption{Variation of optimal model parameters with mean flow speed $v$ (A-E)  
and velocity correlation length $L_{\rm v}$. Data points in (A-E) have $L_{\rm v}\simeq40 \mu m$ and data points in (F-J) have  $v\simeq42\mu m/s$.
The numerical values of all these parameters are listed in {\it SI Appendix, Table~S1}.
Dashed lines in A to I are linear fits (see text); dashed line in J is a fit of $R_{\rm a}\sim\sqrt{\L_{\rm v}}$ with a $R^{2}=0.925$.}
\label{fig7}
\end{figure}

\section*{Discussion}

To summarize, we presented a systematic study of collective motion and defect properties in a dense, 
wet active nematic system composed of filamentous bacteria, and introduced a minimal microscopic model to account for our experiments. 
We have shown that using both orientation and velocity measurements enables to determine a unique, optimal set of parameter values at which
our Vicsek-style model for active suspensions accounts quantitatively for many if not all quantities that one can extract from experimental data.
Because the collective dynamics of our bacterial active nematics is always chaotic, we have used topological defects to
estimate these optimal parameter values. As a matter of fact, it is sufficient to use a small subset of the various quantities we measured 
to determine all optimal parameter values, after which the remaining subset is ``automatically'' matched too. 
The existence of a unique optimum at which matching is nearly perfect constitutes, in retrospect, evidence of the quality of our model.

Thanks to quantitative match at a remarkable level of detail, 
the interplay between experiments and model provides new information and a deeper understanding of our system.
This is in particular the case for the dynamics and structure of topological defects.
Fig.~5C demonstrates that  $+\frac{1}{2}$ defects move approximately twice as fast as the background flow. 
This acceleration can be explained by the local flow field (Fig.~4C), 
which shows two vortices above and below the strong jet advecting the defects. The shape of defects is essentially governed by the vorticity coupling constant $C_{\rm v}$ 
(the orientation field around $+\frac{1}{2}$ defects changes from from arrow-like to mushroom-like shape 
when increasing $C_{\rm v}$ \cite{ZhangRui2017,Kumar2018}). This cannot be seen in our experiments,
in which $C_{\rm v}$ is essentially constant (Fig.~7C,H) but is shown by simulations of our quantitatively faithful model ({\it SI Appendix, Fig.~S5}).
Thus, the vortices can destabilize orientational order ahead of the core, causing the defect to move faster than the background flow. The accelerating effect weakens when particles are less sensitive to flow vorticity or when nematic interaction becomes stronger, as shown in simulation ({\it SI Appendix, Fig.~S4C}) and experiments (Fig.~5CD). 

The \textit{averaged} orientation and velocity fields around $-\frac{1}{2}$ defects (Fig.~4B,D) show an approximate 3-fold rotational symmetry, 
which is consistent with the conventional, equilibrium picture: this symmetry implies that 
active and elastic stresses are balanced around the defect core, and that $-\frac{1}{2}$ defects are passive particles advected by the background flow. 
However, our experiments and simulations indicate that $-\frac{1}{2}$ defects possess a small but significant velocity in the fluid frame $\Delta {\bf v}^-$ (Fig.~3E,I,G,K,5D). Moreover, instantaneous fields around $-\frac{1}{2}$ defects often deviate significantly from 3-fold symmetry, see {\it SI Appendix, Fig.~S9}. 
Such deviations break stress balance around the core and give $-\frac{1}{2}$ defects their velocity over the background flow.
Our data ({\it SI Appendix, Fig.~S9 and S10}) indeed show that the degree of deviation from 3-fold symmetry correlates with this velocity.

Our work also explains the multiple effects of cell length (under the influence of Cephalexin). Cell length directly, and not surprisingly,
governs all lengthscales in our system, and does so nearly identically (Figs.~2E, 5AB, 7F-J). 
More surprising is the observation that the relative speed of defects decreases with cell length (Fig.~5CD) and that the
strain coupling constant $C_{\rm s}$ decreases for long cells (Fig.~7I).

These findings are just a subset of all those illustrating how, thanks to the quantitative modeling, one can not only determine
key effective parameters (such as the strength of flagella or the effective viscosity of our suspension) but also ``read'' important
physical mechanisms from observing how model parameters change in experiments or are changed in simulations.

Our data-driven quantitative matching was made possible thanks to the relative simplicity of our Vicsek-style model: 
even though it deals with wet active suspensions, it possesses a relatively small number of parameters, and is numerically efficient. 
Treating near-field interactions only effectively, it is also versatile, and we believe the same approach can be applied to other active suspensions and extended to include other effects, such as external field and polar order.

The simplicity of our model should also allow to derive continuous, hydrodynamic equations. 
Works on hydrodynamic theories of wet active nematics abound, but they typically lack a direct connection to microscopic mechanisms. Thus deriving a faithful hydrodynamic theory from our quantitatively-valid model is a very promising step.
That would in particular allow to estimate how far our active nematics deviates from elastic theory predictions, something hinted by
the structure of defects (Fig.~4EF). This is the topic of ongoing work at the end of which, we hope, we will be able to build a fully quantitative link between experiments and hydrodynamic theory.

\matmethods{\subsection*{Bacteria strain and Colony Growth}

We use wild type \textit{S. marcescens} strain ATCC 274 labeled with
green fluorescent protein p15A-eGFP. Bacteria colonies are grown on
soft ($0.5\%$) Difco agar plate containing $2.5\%$ Luria Broth (Sigma).
We mix Cephalexin with molten agar at $70^{\circ}C$. We then pour 40 milliliters
of molten agar  into a 15-cm-diameter Petri dish, which is then
dried with a lid on for 16 hours ($25^{\circ}C$ and $50\%$ humidity). 
About $10\text{ }\mu L$ of overnight bacteria culture is then inoculated on
the agar. The inoculated plates are dried for another $15\text{ min}$
without a lid, then stored in an incubator at $30^{\circ}C$ and $90\%$ humidity.

\subsection*{Imaging procedure}

After a growth time of $8-9$ hours, collective motion is observed for as long as 2 hours near the expanding edge of a colony,
in an active region about $1\text{ mm}$ wide. 
The colony expansion speed is approximately $2 \mu m/s$, i.e. much smaller than the measured bacteria flow speed. Thus its influence on bacteria velocity measurement can be neglected.
We capture bacteria motion in the central
part ( $277\times277\mu m^{2}$) of this active region through
a $40\times$objective (Nikon S plan Fluor).
\textit{S. marcescens} colonies quickly change from mono-layer to 3 layer within $100 \mu m$ from the swarming edge, thus the thickness of swarming cells are constant in the observation region.
A Nikon MBE45510 filter
cube (excitation $470/40nm$, emission $525/50nm$) is used for fluorescent
imaging. Images are acquired by a high-speed camera (Basler acA2040-180km)
at $100\text{ frame}/s$ for $30s$, during which bacteria motility remains unchanged.
Bacteria form an immobile film in the central part of the colony. We take videos far enough from this immobile region .}

\showmatmethods{} 

\acknow{X.-q.S., H.C. and H.P.Z. acknowledge financial support of National Natural Science Foundation of China ( Grants No. 11422427, 11774222 to H.P.Z., Grant No. 11635002 to X.-q.S. and H.C., Grants No. 11474210 and No. 11674236 to X.-q.S.). H.P.Z. thanks Program for Professor of Special Appointment at Shanghai Institutions of Higher Learning (No. GZ2016004). H.C. thanks the French ANR project ``Baccterns''. H.P.Z. thanks Julia Yeomans for useful discussions at the initial stage of this work.}

\showacknow{} 

\bibliography{BacteriaMotionPhysics}

\end{document}